# Detection of two-dimensional small polarons at oxide interfaces by optical spectroscopy

Chi Sin Tang,[1,2,3] Shengwei Zeng,[4] Jing Wu,[3] Shunfeng Chen,[1] Muhammad A. Naradipa,[4] Dongsheng Song,[5] M. V. Milošević,[6] Ping Yang,[2] Caozheng Diao,[2] Jun Zhou,[3] Stephen J. Pennycook,[7,9] Mark B.H. Breese,[2,4] Chuanbing Cai,[1] Thirumalai Venkatesan,[10] Ariando Ariando,[4] Ming Yang,[8,*] Andrew T.S. Wee,[4,9,*] Xinmao Yin [1,*]

**Affiliations:**

[1]Shanghai Key Laboratory of High Temperature Superconductors, Department of Physics, Shanghai University, Shanghai 200444, China

[2]Singapore Synchrotron Light Source (SSLS), National University of Singapore, Singapore 117603

[3]Institute of Materials Research and Engineering, A∗STAR (Agency for Science, Technology and Research), 2 Fusionopolis Way, Singapore, 138634 Singapore

[4]Department of Physics, Faculty of Science, National University of Singapore, Singapore 117542

[5]Institute of Physical Science and Information Technology, Anhui University, Hefei 230601, China

[6]Departement Fysica, Universiteit Antwerpen, Groenenborgerlaan 171, B-2020 Antwerpen, Belgium

[7]Department of Materials Science and Engineering, National University of Singapore, Singapore, 117575, Singapore

[8]Department of Applied Physics, The Hong Kong Polytechnic University, Kowloon, Hong Kong, China

[9]Centre for Advanced 2D Materials and Graphene Research, National University of Singapore, Singapore 117546

[10]Center for Quantum Research and Technology, University of Oklahoma, Norman, Oklahoma 73019, USA

*Correspondence to: kevin.m.yang@polyu.edu.hk (M.Y.), phyweets@nus.edu.sg (A.T.S.W)，yinxinmao@shu.edu.cn (X.Y.)

Two-dimensional (2D) perovskite oxide interfaces are ideal systems to uncover diverse emergent properties such as the arising polaronic properties from short-range charge-lattice interactions. Thus, a technique to detect this quasiparticle phenomena at the buried interface is highly coveted. Here, we report the observation of 2D small-polarons at the LaAlO$_3$/SrTiO$_3$ (LAO/STO) conducting interface using high-resolution spectroscopic ellipsometry. First-principles investigations shows that interfacial electron-lattice coupling mediated by the longitudinal phonon mode facilitates the formation of these polarons. This study resolves the longstanding question by attributing the formation of interfacial 2D small polarons to the significant mismatch between experimentally measured interfacial carrier density and theoretical values. Our study sheds light on the complexity of broken periodic lattice-induced quasi-particle effects and its relationship with exotic phenomena at complex oxide interfaces. Meanwhile, this work establishes spectroscopic ellipsometry as a useful technique to detect and locate optical evidence of polaronic states and other emerging quantum properties at the buried interface.

**Introduction**

The emergence of quasiparticles due to the interplay between electronic and lattice degrees of freedom in strongly-correlated systems is the cornerstone of multiple fundamental phenomena including transport processes, colossal magnetoresistance[1] and high-temperature superconductivity[2]. Polarons are an ideal example where motion of mobile charges are retarded due to strong charge-lattice interaction [3]. Though well-established in 3D systems[4], polaronic formation and modifications to their properties by reduced dimensionality in 2D systems remains challenging due to the incomplete understanding of polaron self-trapping in anisotropic structures[5]. Nevertheless, the key role of strongly-bound electronic polarons and bipolarons which possess lower-temperature coherence could hold the key in understanding many-body phenomena such as itinerant ferromagnetism[6] and unconventional Cooper pairing in superconductivity[7, 8]. Such prospects are tantalizingly attractive especially with reports of 2D polaronic behaviour[4, 5] alongside their causality with 2D superconductivity[9]. The effective identification of polaronic activities and the underlying mechanisms leading to their formation play a critical role in assessing the distinguish the short- and long-range transport characteristics and sheds new light on the nature of quasiparticle interactions in diverse material systems[4]. Such knowledge holds immediate relevance especially in the functionalization of novel quantum systems in applications over a wide range of domains related to electronic transport[10, 11], photocatalysis[12] and energy storage devices[13].

Recent reports of 2D perovskite oxide superconductivity at the $LaAlO_3/KTaO_3$ interface have renewed fresh interests in the emergence of anomalous quantum metallic states[14, 15]. The quintessential $LaAlO_3/SrTiO_3$ (LAO/STO) conducting interface remains the ideal candidate for such investigation especially with the versatility and capacity to effectively manipulate its charge, spin, lattice, and orbital degrees-of-freedom at the nanometer-thickness interface[16]. Furthermore, insights to the interfacial many-body charge dynamics in the region ~10 meV within the Fermi level is pivotal to explain the severe discrepancy between the theoretically predicted 0.5 electrons per unit cell ($e^-$/uc) charge transfer by multiple models[10, 17, 18], as compared to a meagre 0.05 $e^-$/uc elucidated by experiments[19-21].

Here, we make the observation of small-polarons at the LAO/STO interface using high-resolution spectroscopic ellipsometry, a highly sensitive non-destructive photon-in photon-out optical technique with a spectral resolution of 0.02 eV which, with its full polarization and symmetry features, can explore the role of electron-phonon coupling and effectively resolve the anisotropic properties of the LAO/STO system and extract the optical properties of its interfacial 2D electron gas. This is in light of the progress in the experimental techniques to characterize polaronic responses, that while there were initially no evidence of sufficiently large electron-phonon coupling to induce small polaron formation in Nb-doped STO[22], a subsequent experimental study then suggested polaronic correlations in the LAO/STO system which possesses strongly frequency-dependent mobility similar to that of Nb-doped STO[23]. Even though significant breakthrough was then made with the observation of large polaron at the LAO/STO interface using angle resolved photoemission spectroscopy (ARPES)[10], it can only study the interfacial properties in the vicinity of the fermi level. The unique properties of spectroscopic ellipsometry allows one to study the electronic properties and quasiparticle dynamics of systems into the near infrared, visible and ultraviolet regime[24-27] that provide evidence in identifying the optical response of the small polarons presented at the 2D LAO/STO interface. First-principles calculations further suggest the strong coupling between the interfacial electrons and the Ti-lattice as the key mechanism leading to the formation of the localized small polarons which is 2D in nature. Based on an integrated analysis of our experimental results and findings from the first-principle studies, the hard longitudinal optical phonon mode, LO3, which has previously been identified as the key mediator in the formation of large polarons[10, 13], has also been determined to play a critical role in the formation of the 2D small-polarons. Importantly, by showing that ~50% of the interfacial charges couple strongly with the Ti lattice sites to form highly localized 2D small polarons [Fig. 1(a)], this study effectively resolved the longstanding question why the experimentally-measured interfacial carrier density is significantly lower than theoretically predicted values[10, 17, 18]. In light of how quasiparticle dynamics governs superconductivity at perovskite oxide interfaces, our study further highlights the strong interactions between excess interfacial electrons and Ti lattices that results in the polaronic states where lattice distortion invariably breaks the periodic lattice symmetry[4]. This study is therefore potentially analogous to other heterostructure systems such as that of magic-angle twisted bilayer graphene where their superconductive states could be attributed to the many-body correlations induced by broken periodic lattice symmetry[28-32]. With the escalating challenge to characterize emergent quantum orders in complex low-dimensional systems, this study further highlights spectroscopic ellipsometry as the premier experimental methodology to unambiguously identify and analyse anisotropic quasiparticle dynamics and other emergent order-parameter nanostructures at the buried complex quantum interfaces.

## Results

### Sample Synthesis and Characterization

Three high-quality single crystalline 8 u.c. LaAlO$_3$/SrTiO$_3$ labelled *N1*-LAO/STO, *N2*-LAO/STO, and *N3*-LAO/STO, respectively, are synthesized with increasing degree of oxygen vacancies, thereby resulting in different carrier concentrations – $9.4 \times 10^{12}$ cm$^{-2}$ (*N1*-LAO/STO), $1.0 \times 10^{14}$ cm$^{-2}$ (*N2*-LAO/STO), and $4.5 \times 10^{16}$ cm$^{-2}$ (*N3*-LAO/STO) [Fig. S1] (details in Section I.B, supplementary material) as further confirmed based on synchrotron-based X-ray absorption spectroscopy [Fig. 1] (details in Section I.B, supplementary material). The quality of the LAO/STO interfaces has been confirmed using a series of experimental techniques including atomic force microscopy (AFM) [Fig. 1(a)], high angle annular dark field STEM (HAADF–STEM) [Fig. 1(b)] and high-resolution X-ray diffraction [Figs. S2–S4] which suggest atomically flat terrace structures, high-quality morphology, and crystallinity.

### Temperature-dependent Spectroscopic Ellipsometry

Temperature-dependent optical characterization of *N1*-LAO/STO is performed using Spectroscopic Ellipsometry with the complex dielectric function, $\varepsilon(\omega) = \varepsilon_1(\omega) + i\varepsilon_2(\omega)$ [Fig. 2(a)] (details in Section II.D, supplementary material). The fitted results of both the $\varepsilon_1$ and $\varepsilon_2$ spectra show Kramers-Kronig consistency and they are in good agreement with previous experimental studies in regions above 1 eV[33]. The optical features in the photon energy region of ~2–3 eV are likely the result of interband transitions. As seen thereafter in the electronic band structure [Fig. 4(g)], the feature at ~3.0 eV can be ascribed to the valence to conduction band edge transition while the features in between 2—3 eV are likely due to optical transitions between the valence band and polaron band (to be discussed thereafter). The Drude response below 1 eV suggests that it is a conducting system ascribed to the interfacial 2D delocalized electrons. Interestingly, a previously unidentified prominent mid-gap peak feature is observed in the near-infrared regime at ~0.66 eV (annotated by arrow) throughout all temperature which has not been observed or scrutinized previously. With the repeated characterization of similar LAO/STO interface synthesized under the same experimental conditions, this feature persists. Thus, confirming that this is not an optical artefact that arises from the optical measurements. Besides, with the focus of this study on the optical features in the near-IR photon energy region, it allows for the detailed scrutiny to account for the origin of this previously unaccounted for optical feature. With minimal oxygen defect from *N1*-LAO/STO and having previously confirmed the high-quality and crystallinity of the sample, one can rule out the contribution of oxygen vacancies and defects states (Discussion eliminating other factors to be discussed thereafter). Instead, it bears resemblance to the optical absorption feature of small polaron due to its asymmetry[34]. As discussed in greater detail thereafter, this feature is attributed to the interfacial 2D small polarons.

The $\varepsilon_2$ spectra of *N1*-LAO/STO is further converted to optical conductivity, $\sigma_1$, using Supplementary Equation 11 (details in Section III.A. supplementary material) for further polaron analysis [Fig. 2(b)]. While the previously observed Drude response falls below the spectral range, the polaron peak remains prominent at ~0.82–0.98 eV where its asymmetric shape characteristic of small polaron remains[34]. The prominent nature of this optical peak clearly indicates that it is a standalone optical feature independent of any other features present in the optical spectra. Since the small polarons optical features arise due to electron hopping between in-plane neighbouring Ti-sites upon activation, the peaks can thus be modelled using the Bryksin small-polaron model to analyze its temperature-dependent properties[35]. Specifically, parameters including the polaron hopping energy, $E_a$, bandwidth, $\Gamma$, and phonon energy, $E_{LO}$, can be derived (details in Section III.A, supplementary material).

The optical polaron responses is very compatible with the theoretical Bryksin small-polaron model for both *N1*- and *N2*-LAO/STO interfaces, respectively, throughout the entire temperature range as seen in the zoom in stacked $\sigma_1$ spectra in [Figs. 2(c) and 2(f)] (further temperature-dependent analyses to be discussed in Fig. S3 and Section III.B, supplementary material). This indicates that the near-infrared responses in both *N1*- and *N2*-LAO/STO interfaces are of small-polaron origin. Besides, they are notably distinct from the previously identified interfacial large polarons[10] as confirmed in the computational studies presented thereafter.

We analyze the temperature-dependent $\varepsilon_1$ and $\varepsilon_2$ spectra, respectively, of *N2*-LAO/STO [Fig. 2(d)] to investigate how increasing interfacial carrier concentration affects the modifies the polaronic response. Once again, while the optical spectra of both the *N1*- and *N2*-LAO/STO interfaces are Kramers-Kronig consistent and display similar optical features with previous reports[33], the Drude response for *N2*-LAO/STO due to higher interfacial carrier concentration. Notably, the mid-gap previously attributed to the small polaron response persists at the *N2*-LAO/STO interface across all temperature at a redshifted position of ~0.42 eV. Similar to the previously characterized *N1*-LAO/STO, this optical feature persists present for other synthesized LAO/STO interfaces possessing the same conditions as N2-LAO/STO. thereby confirming once again that this is not a mere optical artefact. Related to the polaron activation energy $E_a$, this redshift in the polaron peak is consistent with our theoretical study where LAO/STO interfaces with higher interfacial charge concentration tend to have a lower $E_a$, as discussed in greater detail thereafter. Based on the Bryksin analysis (Section III.A, supplementary material), while the Drude response is red-shifted below the spectral range, the polaron peak remains prominent [Fig. 2(e)] and they display high consistency between the experimental results and theoretical fitting throughout the entire measured temperature range as seen in the temperature-dependent stacked $\sigma_1$ spectra belonging to *N2*-LAO/STO in [Fig. 2(f)] where the polaron activation energy, $E_a$, is significantly lower (Fig. S3 and Section III.B, supplementary material).

Compared to *N1* and *N2*-LAO/STO, *N3*-LAO/STO shows a marked increase in temperature-dependence especially for the $\varepsilon_1$ spectra while the $\varepsilon_2$ spectra undergoes a significant change in the low-energy region where the previously observed near-infrared polaron peak is further redshifted to photon energy below the instrument spectral limit [Fig. 3(a)]. Hence, analysis of the small polaron response will be restricted to the *N1*- and *N2*-LAO/STO interfaces (details in Section III.B, supplementary material).

**Analyzing the Small Polaron Parameters**

Using Supplementary equation 9b in the supplementary material to perform curve fitting for the respective $\sigma_1$ spectra in the photon energy region where the polaron peak is located (stacked $\sigma_1$ spectra of both *N1*- and *N2*-LAO/STO interfaces displayed in Figs. 2(c) and 2(f), respectively), the hopping energy, $E_a$, polaron band width, $\Gamma$, and phonon energy, $E_{LO}$, as displayed in Figs. 3(b)–3(d) respectively, of the 2D interfacial polarons at the *N1*- and *N2*-LAO/STO interfaces are elucidated. $E_a$ of *N2*-LAO/STO falls significantly lower to the region between ~0.11—0.12 eV compared to *N1*-LAO/STO's ~0.24—0.25 eV. Between 77 and 300K, temperature-dependence of $E_a$ of both *N1*- and *N2*-LAO/STO follows similar trend with a slight but apparent increase with rising temperature as it becomes easier for the electron to move from one Ti-site to the next[34]. The distinct reduction in $E_a$ of *N2*-LAO/STO compared to *N1*-LAO/STO may be partially attributed to the increase in overlapping potential wells due to higher oxygen vacancies at the interface that allows for a greater ease of interstitial charge hopping. As scrutinized in greater detail thereafter, theoretical calculations quantitively show that the polaron state of the LAO/STO interface with higher 2D excess electrons is formed at an energy position nearer to the conduction band edge along the $\Gamma$-point of the band structure. This quantitively accounts for why the polaron activation energy, $E_a$, for the more conducting $N2$-LAO/STO is lower than that of $N1$-LAO/STO which possesses lower interfacial charge concentration (see [Fig. 4(g)]). Comparison of $E_a$ between $N1$-LAO/STO and $N2$-LAO/STO further suggests that an increase in interfacial charge density has a greater influence on the interfacial polarons than temperature.

The polaron bandwidth, $\Gamma$, of $N2$-LAO/STO is slightly larger than $N1$-LAO/STO [Fig. 3(c)]. Nevertheless, both display similar temperature-dependent trends with a monotonic width broadening from ~0.039 to ~0.057 eV for $N2$-LAO/STO and relatively smaller increase of ~0.019 to ~0.026 eV for $N1$-LAO/STO. This progressive broadening with rising temperature is consistent with the behaviour of small-polarons which corresponds with the phonon broadening of the local electronic energy levels which in turn progressively broadens these absorption bands[34]. To account for the relatively larger bandwidth belonging to $N2$-LAO/STO as compared to $N1$-LAO/STO, it is first noted that the conduction band of the LAO/STO interface is mainly dominated by multiple overlapping bands belonging to the Ti $d$-orbitals (compare conduction band states of the Ti and O orbitals based on PDOS calculations in Figs. 4(e) and 4(f), respectively). With an increase in the charge concentration at the LAO/STO interface through the introduction of oxygen vacancies, there will invariably be an increase in the number of conduction band states that will be occupied when polaron transitions take place. As a result, one will expect an increase in the distribution of polaron energy as evidenced by the broadening of the polaron bandwidth, $\Gamma$, as observed in our experimental result[35, 36].

**Role of LO3 in the Formation of the Interfacial Small Polarons**

Finally, the $E_{LO}$ of both $N1$- and $N2$-LAO/STO does not have any clear temperature trend [Fig. 3(b)] but their values fluctuate between ~0.102–0.106 eV for $N1$-LAO/STO and ~0.13–0.18 eV for $N2$-LAO/STO. These values are consistent with the hard longitudinal optical phonon, LO3. This phonon mode is attributed to the atomic displacements leading to the breathing distortion of an octahedral cage that surrounds a Ti-site[10]. The role that LO3 plays in mediating electron-phonon coupling is not only restricted to the LAO/STO interface, but also on the surfaces and within the bulk of STO substrates[22, 37-39]. In the case of the LAO/STO interfacial polaron reported in this study, it is consistent with previous studies where there are barely any changes in energy position with temperature[10, 40]. Moreover, as the LO3 phonon falls within the energy range of ~100 meV where our interfacial small polaron are detected, it is the most probable phonon mode that mediates the formation of the interfacial small polarons observed in our study. Besides, as it possesses the highest electron-phonon coupling constant amongst the LO-modes[40-43], it allows for the formation of small polarons particularly facilitated by short-range electron-phonon interaction. The presence of LO3 phonon mode plays an important role in controlling the 2D interfacial carrier mobility[10]. Considering the role that LO3 plays in the interfacial polaron formation, the larger $E_{LO}$ magnitude for *N2*-LAO/STO may be attributed to structural changes due to electrostatic doping by the increase in oxygen vacancies[44] or possible weaker coupling to additional phonon modes. To further account for the slight discrepancies between the theoretical energy position of the LO3 phonon (at ~100 meV)[10] and that which is derived from our data fitting, interfacial electric field, phonon coupling across the LAO/STO interface, propagation lattice distortion across the interface may be likely factors leading to such quantitative differences[45]. Nevertheless, the phonon mode analysis provided in this study and the preceding ones highlights the critical role that electron-phonon interaction plays in the formation of this 2D interfacial polaron. The identification of $E_{LO}$ as the energy position of LO3 phonons is a clear indication of polarons and the identification of polaronic effects[46, 47] while distinguish it from other phenomena. Besides, while the large polarons in LAO/STO have been attributed to the LO3 phonons[10], this analysis conclusively shows that the LO3 phonons also plays the direct role in the formation of the small polarons at the LAO/STO interface.

Overall, the compatibility between the optical features in the respective $\sigma_1$ spectra and the Bryksin small-polaron model alongside the considerably reasonable fitting error (as denoted by the error bars) are indications that these optical responses are indeed signatures of small polaron present at the LAO/STO interface. Besides, the general temperature-dependent trends and asymmetric peak feature of both *N1*- and *N2*-LAO/STO provide further indications of the small polaron dynamics at the LAO/STO interfaces[35]. The temperature-dependent behaviours also allow us to rule out the contribution of large polaron to these optical features[34]. The use of the Bryksin model to verify the small interfacial behaviour at the LAO/STO interface is further confirmed and substantiated via first-principles study as discussed thereafter.

**Further Eliminating other Possible Contributing Factors to the Optical Responses**

Besides ruling out the possibility of defect states contributing to the formation of the near-infrared optical response, we rule out any other possible factors that may lead to the optical features in Fig. 2. The presence of exciton may be omitted because while exciton weakens significantly with rising temperature[48, 49], the near-IR feature detected at the LAO/STO interface has a trend contrary to that of exciton with an intensity increase with temperature.

With the vanishingly small ferromagnetic response contributed by the LAO/STO interface[50], the onset of magnons can also be omitted. Finally, any contribution by plasmon excitation can also be ruled out due to detection of plasmon requires the simultaneous presence of a zero-crossing in the $\varepsilon_1$ spectrum alongside a prominent peak feature in the Loss-function spectrum which are not present in the respective components of the dielectric functions[26, 51].

Therefore, by eliminating these possible contributing factors, the polaronic origin of these mid-IR peaks at the LAO/STO interface can be safely concluded.

**First-principles Calculations Elucidating Small Polaron Properties**

Extensive first-principles studies are conducted to further substantiate our experimental findings on how excess electrons interact with the interfacial Ti ions to form small polarons and account for the missing interfacial electrons. The LAO/STO interface was modelled using a $(LaAlO_3)_{6.5}/(SrTiO_3)_{8.5}$ superlattice [Fig. 4(a)] (details in Section V, supplementary material) with 0.5 $e^-$/uc charge transfer from the LAO layer to the interface. Previous studies have shown that this modelized structure has been widely used to mimic and simulate the LAO/STO interfacial effects[52-55]. Hence, the results yielded by the current study provide important insights to the phenomena observed in real experimental systems. Besides, the polaron formation energy at this charge-transfer density is ~160 meV lower than the non-polaron state [Fig. S8] (details in Section V.B, supplementary material). The interfacial polaron states are therefore more stable than the free-electron states. While charge transfer from the LAO layer penetrates the interface into the STO layer and localize at the Ti lattice sites (blue superimposed partial charge density in the STO layer in [Fig. 4(b)], it is only at the interfacial Ti lattice sites that the localized charge concentration is sufficiently large to form small polarons [Fig. 4(a)]. Small polaron formation is further evidenced by a slight elongation of the in-plane Ti-O bonds by about +0.08 Å relative to Ti-sites where there are no polaron states as seen in the interfacial lattice cross-section [Fig. 4(c)]. This is consistent with the magnitude of lattice deformation due to small polaron formation as reported in other systems such as $TiO_2$ and Nb-doped STO[56-58]. Moreover, as noticed in the top view of interfacial structure superimposed with the partial charge density of the polaron state [Fig. 4(d)], the electron density only localizes at the distorted Ti site and not homogeneously at the others. With the concurrent presence of local lattice deformation as seen in the Ti-O elongation [Fig. 4(c)] and the onset of localized electron density at the affected Ti-lattice sites [Fig. 4(d)], these are clear fingerprints of the formation of small polarons[56]. Collectively, these are evidence of small polaron states with *two-dimensional* character.

The 2D small polarons is further confirmed by the projected density of states (PDOS) at the interfacial TiO$_2$-sublayer. [Fig. 4(e)] shows that strong charge-lattice interaction causes a split in the in-plane Ti-$d_{xy}$ states from the conduction band edge and forms an isolated mid-gap state – another distinctive hallmark of the polaron state – which hybridizes weakly with the nearest in-plane O-$p_y$/$p_y$ states [Fig. 4(f)]. This is yet another signature of a polaron state with two-dimensional feature due to the in-plane hybridized orbitals - consistent with the prominent in-plane Ti-O bond elongation alongside the in-plane polaron charge density distribution state [Figs. 4(a)—4(d)]. Therefore, it can be concluded to be an interfacial *two-dimensional* small polaron. By further modelling the partial charge density for the polaronic band in the energy range between -1 and -2 eV, the integration of the partial charge density leads a value of 0.25 $e^-$/uc – half of the total transferred electron of 0.5 $e^-$/uc. Hence, we deduced that about half of the electrons interact with the interfacial Ti lattice to form the polaron state [Fig. 4(a)] where it is further confirmed by a previous study[10].

In tandem with the experimental study involving LAO/STO interfaces with different charge concentrations, a less conducting LAO/STO interface with 0.4 $e^-$/uc charge transfer is modelled. While its polaron stability is reduced, the interfacial polaron state remains favourable with an energy of 126 meV lower than the non-polaron state [Fig. S8] (details in Section V.B, supplementary material) due to decreased electron-electron repulsion that weakens the excess electrons interaction and the Ti lattice sites[56].

The reduction in carrier density also has strong influence on the electronic properties of the LAO/STO interface with the polaron. Fig. 4(g) compares the band structures of the LAO/STO interfaces with 0.5 $e^-$/uc (black solid lines) and 0.4 $e^-$/uc charge transfer (red dashed lines), respectively. With the conduction band edge of both interfaces aligned at -0.365 eV indicated by the overlapping black and red bands at the $\Gamma$-point, the polaron state for the less conducting system is located ~0.15 eV *lower* than the more conducting interface – denoted by yellow shaded region in [Fig. 4(g)]. This is consistent with the experimental results that compares *N1*-LAO/STO which has a lower interfacial charge concentration than *N2*-LAO/STO [Figs. 2(b) and 2(e), respectively]. The polaron positions – directly related to the polaron activation energy, $E_a$ – of the more conducting *N2*-LAO/STO is lower compared to *N1*-LAO/STO [Fig. 3(b)].

This also explains why the polaron position falls below the instrument spectral range for *N3*-LAO/STO.

**Discussion and Conclusion**

The combination of experimental results and theoretical calculations provides clear evidence of the onset of 2D small polarons at the LAO/STO interface. This study is important in understanding how quasiparticle dynamics governs superconductivity at perovskite oxides heterointerfaces and is potentially analogous to other heterostructure systems especially in twisted bilayer graphene[30, 59]. The strong interactions between excess interfacial electrons and Ti lattices result in the polaronic states where lattice distortion invariably breaks the periodic lattice symmetry[4, 60]. Meanwhile, our calculations show that the formation of small polarons due to considerable loss in electron kinetic energy. This in turn leads to in a significant reduction in polaron bandwidth to ~0.5 eV compared to ~1.2 eV for free electron states. This band flattening strengthens the electron-electron correlations and is key to the emergence of the interfacial superconductivity[61]. This superconductive mechanism in perovskite oxide interfaces resembles the superconductive properties in magic-angle twisted bilayer graphene with an interlayer misalignment of $1.1°$[29, 30]. Specifically, flat electronic bands are present at certain interlayer twist angles[29, 30]. The formation of the Moiré patterns breaks the original periodic structure, and the inhomogeneity of the electron systems drastically reduces the Fermi velocity. This significantly modifies the electronic properties where flat bands are form at the Fermi level which, in turn, leads to superconductivity[62]. Similar to the interfacial superconductivity of oxide heterostructures, the emergence of strong electron-electron correlations because of broken periodic lattice symmetry holds the key to the onset of superconductivity in magic angle twisted bilayer graphene[28] (details in Section VI, supplementary material).

The formation of 2D small polaron at the LAO/STO interface has led to the localization of 50% of the interfacial excess electrons. Notably, the interfacial electron density and lattice distortion are pivotal in dictating the small polaron dynamics and it holds important implications on how quasiparticle dynamics mediates superconductivity in complex heterointerfaces including perovskite oxides and magic-angle twisted bilayer graphene. It further highlights spectroscopic ellipsometry as the ideal experimental technique to unambiguously identify and characterize anisotropic quasiparticle dynamics and other emergent order-parameter nanostructures at the buried complex quantum interfaces and other low-dimensional heterostructure systems.

**Acknowledgements:**

This work was supported in part by the Strategic Priority Research Program of the Chinese Academy of Sciences, Grant No. XDB25000000, National Natural Science Foundation (52172271), the National Key R&D Program of China No. 2022YFE03150200. This research is also supported by the Agency for Science, Technology, and Research (A*STAR) under its Advanced Manufacturing and Engineering (AME) Individual Research Grant (IRG) (A1983c0034) and the National Research Foundation, Singapore, under the Competitive Research Programs (CRP Grant No. NRF-CRP15-2015-01). C. S. T. acknowledges the support from the NUS Emerging Scientist Fellowship. J. W. acknowledge the Advanced Manufacturing and Engineering Young Individual Research Grant (AME YIRG Grant No.: A2084c170). J. Z. and M. Y. would like to acknowledge Singapore MOE Tier 2 grant (MOE2019-T2-2-30) and computing resource provided by the Centre for Advanced 2D Materials and Graphene Research at National University of Singapore, and National Supercomputing Centre of Singapore. The authors would like to acknowledge the Singapore Synchrotron Light Source for providing the facility necessary for conducting the research. The Laboratory is a National Research Infrastructure under the National Research Foundation, Singapore. Any opinions, findings and conclusions or recommendations expressed in this material are those of the author(s) and do not reflect the views of National Research Foundation, Singapore.

**References:**

1. N. Mannella, W. L. Yang, X. J. Zhou, H. Zheng, J. F. Mitchell, J. Zaanen, T. P. Devereaux, N. Nagaosa, Z. Hussain and Z. X. Shen, Nature **438** (7067), 474-478 (2005).
2. P. A. Lee, N. Nagaosa and X.-G. Wen, Reviews of Modern Physics **78** (1), 17-85 (2006).
3. T. Holstein, Annals of Physics **8** (3), 343-389 (1959).
4. C. Franchini, M. Reticcioli, M. Setvin and U. Diebold, Nature Reviews Materials **6** (7), 560-586 (2021).
5. K. P. McKenna, M. J. Wolf, A. L. Shluger, S. Lany and A. Zunger, Physical Review Letters **108** (11), 116403 (2012).
6. G. Valtolina, F. Scazza, A. Amico, A. Burchianti, A. Recati, T. Enss, M. Inguscio, M. Zaccanti and G. Roati, Nature Physics **13** (7), 704-709 (2017).
7. M. Danilov, E. G. C. P. van Loon, S. Brener, S. Iskakov, M. I. Katsnelson and A. I. Lichtenstein, npj Quantum Materials **7** (1), 50 (2022).
8. K. Jin, N. P. Butch, K. Kirshenbaum, J. Paglione and R. L. Greene, Nature **476** (7358), 73-75 (2011).
9. M. Kang, S. W. Jung, W. J. Shin, Y. Sohn, S. H. Ryu, T. K. Kim, M. Hoesch and K. S. Kim, Nature Materials **17** (8), 676-680 (2018).
10. C. Cancellieri, A. S. Mishchenko, U. Aschauer, A. Filippetti, C. Faber, O. S. Barišić, V. A. Rogalev, T. Schmitt, N. Nagaosa and V. N. Strocov, Nature Communications **7** (1), 10386 (2016).
11. J. C. Garcia, M. Nolan and N. A. Deskins, The Journal of Chemical Physics **142** (2), 024708 (2015).
12. D. Ghosh, E. Welch, A. J. Neukirch, A. Zakhidov and S. Tretiak, The Journal of Physical Chemistry Letters **11** (9), 3271-3286 (2020).
13. W. Zheng, B. Sun, D. Li, S. M. Gali, H. Zhang, S. Fu, L. Di Virgilio, Z. Li, S. Yang, S. Zhou, D. Beljonne, M. Yu, X. Feng, H. I. Wang and M. Bonn, Nature Physics **18** (5), 544-550 (2022).
14. C. Liu, X. Yan, D. Jin, Y. Ma, H.-W. Hsiao, Y. Lin, M. Bretz-Sullivan Terence, X. Zhou, J. Pearson, B. Fisher, J. S. Jiang, W. Han, J.-M. Zuo, J. Wen, D. Fong Dillon, J. Sun, H. Zhou and A. Bhattacharya, Science **371** (6530), 716-721 (2021).

15. Z. Chen, Y. Liu, H. Zhang, Z. Liu, H. Tian, Y. Sun, M. Zhang, Y. Zhou, J. Sun and Y. Xie, Science **372** (6543), 721-724 (2021).
16. H. Y. Hwang, Y. Iwasa, M. Kawasaki, B. Keimer, N. Nagaosa and Y. Tokura, Nature Materials **11** (2), 103-113 (2012).
17. N. Nakagawa, H. Y. Hwang and D. A. Muller, Nature Materials **5** (3), 204-209 (2006).
18. L. Yu and A. Zunger, Nature Communications **5** (1), 5118 (2014).
19. S. Thiel, G. Hammerl, A. Schmehl, C. W. Schneider and J. Mannhart, Science **313** (5795), 1942-1945 (2006).
20. A. Brinkman, M. Huijben, M. van Zalk, J. Huijben, U. Zeitler, J. C. Maan, W. G. van der Wiel, G. Rijnders, D. H. A. Blank and H. Hilgenkamp, Nature Materials **6** (7), 493-496 (2007).
21. N. Reyren, S. Thiel, A. D. Caviglia, L. F. Kourkoutis, G. Hammerl, C. Richter, C. W. Schneider, T. Kopp, A. S. Rüetschi, D. Jaccard, M. Gabay, D. A. Muller, J. M. Triscone and J. Mannhart, Science **317** (5842), 1196-1199 (2007).
22. J. L. M. van Mechelen, D. van der Marel, C. Grimaldi, A. B. Kuzmenko, N. P. Armitage, N. Reyren, H. Hagemann and I. I. Mazin, Physical Review Letters **100** (22), 226403 (2008).
23. A. Dubroka, M. Rössle, K. W. Kim, V. K. Malik, L. Schultz, S. Thiel, C. W. Schneider, J. Mannhart, G. Herranz, O. Copie, M. Bibes, A. Barthélémy and C. Bernhard, Physical Review Letters **104** (15), 156807 (2010).
24. X. Yin, Q. Wang, L. Cao, C. S. Tang, X. Luo, Y. Zheng, L. M. Wong, S. J. Wang, S. Y. Quek, W. Zhang, A. Rusydi and A. T. S. Wee, Nature Communications **8** (1), 486 (2017).
25. X. Yin, S. Zeng, T. Das, G. Baskaran, T. C. Asmara, I. Santoso, X. Yu, C. Diao, P. Yang, M. B. H. Breese, T. Venkatesan, H. Lin, Ariando and A. Rusydi, Physical Review Letters **116** (19), 197002 (2016).
26. X. Yin, C. S. Tang, S. Zeng, T. C. Asmara, P. Yang, M. A. Naradipa, P. E. Trevisanutto, T. Shirakawa, B. H. Kim, S. Yunoki, M. B. H. Breese, T. Venkatesan, A. T. S. Wee, A. Ariando and A. Rusydi, ACS Photonics **6** (12), 3281-3289 (2019).
27. C. S. Tang, X. Yin, S. Zeng, J. Wu, M. Yang, P. Yang, C. Diao, Y. P. Feng, M. B. H. Breese, E. E. M. Chia, T. Venkatesan, M. Chhowalla, A. Ariando, A. Rusydi and A. T. S. Wee, Advanced Materials **32** (34), 2000153 (2020).
28. Y. Xie, B. Lian, B. Jäck, X. Liu, C.-L. Chiu, K. Watanabe, T. Taniguchi, B. A. Bernevig and A. Yazdani, Nature **572** (7767), 101-105 (2019).
29. Y. Cao, V. Fatemi, A. Demir, S. Fang, S. L. Tomarken, J. Y. Luo, J. D. Sanchez-Yamagishi, K. Watanabe, T. Taniguchi, E. Kaxiras, R. C. Ashoori and P. Jarillo-Herrero, Nature **556** (7699), 80-84 (2018).
30. Y. Cao, V. Fatemi, S. Fang, K. Watanabe, T. Taniguchi, E. Kaxiras and P. Jarillo-Herrero, Nature **556** (7699), 43-50 (2018).
31. A. G. Swartz, H. Inoue, T. A. Merz, Y. Hikita, S. Raghu, T. P. Devereaux, S. Johnston and H. Y. Hwang, Proceedings of the National Academy of Sciences **115** (7), 1475 (2018).
32. A. D. Caviglia, S. Gariglio, N. Reyren, D. Jaccard, T. Schneider, M. Gabay, S. Thiel, G. Hammerl, J. Mannhart and J. M. Triscone, Nature **456** (7222), 624-627 (2008).
33. T. C. Asmara, A. Annadi, I. Santoso, P. K. Gogoi, A. Kotlov, H. M. Omer, M. Motapothula, M. B. H. Breese, M. Rübhausen, T. Venkatesan, Ariando and A. Rusydi, Nature Communications **5** (1), 3663 (2014).
34. D. Emin, Physical Review B **48** (18), 13691-13702 (1993).
35. C. A. Triana, C. G. Granqvist and G. A. Niklasson, Journal of Applied Physics **119** (1), 015701 (2016).
36. C. A. Triana, C. G. Granqvist and G. A. Niklasson, Journal of Applied Physics **118** (2), 024901 (2015).
37. Y. J. Chang, A. Bostwick, Y. S. Kim, K. Horn and E. Rotenberg, Physical Review B **81** (23), 235109 (2010).
38. Z. Wang, S. McKeown Walker, A. Tamai, Y. Wang, Z. Ristic, F. Y. Bruno, A. de la Torre, S. Riccò, N. C. Plumb, M. Shi, P. Hlawenka, J. Sánchez-Barriga, A. Varykhalov, T. K. Kim, M. Hoesch, P. D. C. King,

W. Meevasana, U. Diebold, J. Mesot, B. Moritz, T. P. Devereaux, M. Radovic and F. Baumberger, Nature Materials **15** (8), 835-839 (2016).
39. C. Chen, J. Avila, E. Frantzeskakis, A. Levy and M. C. Asensio, Nature Communications **6** (1), 8585 (2015).
40. W. Meevasana, X. J. Zhou, B. Moritz, C. C. Chen, R. H. He, S. I. Fujimori, D. H. Lu, S. K. Mo, R. G. Moore, F. Baumberger, T. P. Devereaux, D. van der Marel, N. Nagaosa, J. Zaanen and Z. X. Shen, New Journal of Physics **12** (2), 023004 (2010).
41. G. Verbist, F. M. Peeters and J. T. Devreese, Ferroelectrics **130** (1), 27-34 (1992).
42. N. Choudhury, E. J. Walter, A. I. Kolesnikov and C.-K. Loong, Physical Review B **77** (13), 134111 (2008).
43. M. Cardona, Physical Review **140** (2A), A651-A655 (1965).
44. D. Doennig and R. Pentcheva, Scientific Reports **5** (1), 7909 (2015).
45. C. L. Jia, S. B. Mi, M. Faley, U. Poppe, J. Schubert and K. Urban, Physical Review B **79** (8), 081405 (2009).
46. V. N. Strocov, C. Cancellieri and A. S. Mishchenko, in *Spectroscopy of Complex Oxide Interfaces: Photoemission and Related Spectroscopies*, edited by C. Cancellieri and V. N. Strocov (Springer International Publishing, Cham, 2018), pp. 107-151.
47. C. Verdi, F. Caruso and F. Giustino, Nature Communications **8** (1), 15769 (2017).
48. F. Cadiz, E. Courtade, C. Robert, G. Wang, Y. Shen, H. Cai, T. Taniguchi, K. Watanabe, H. Carrere, D. Lagarde, M. Manca, T. Amand, P. Renucci, S. Tongay, X. Marie and B. Urbaszek, Physical Review X **7** (2), 021026 (2017).
49. K. He, N. Kumar, L. Zhao, Z. Wang, K. F. Mak, H. Zhao and J. Shan, Physical Review Letters **113** (2), 026803 (2014).
50. A. Fête, S. Gariglio, C. Berthod, D. Li, D. Stornaiuolo, M. Gabay and J. M. Triscone, New Journal of Physics **16** (11), 112002 (2014).
51. D. Pines, Reviews of Modern Physics **28** (3), 184-198 (1956).
52. R. Pentcheva and W. E. Pickett, Physical Review Letters **102** (10), 107602 (2009).
53. Z. Zhong and P. J. Kelly, Europhysics Letters **84** (2), 27001 (2008).
54. Z. S. Popović, S. Satpathy and R. M. Martin, Physical Review Letters **101** (25), 256801 (2008).
55. H. Banerjee, S. Banerjee, M. Randeria and T. Saha-Dasgupta, Scientific Reports **5** (1), 18647 (2015).
56. X. Hao, Z. Wang, M. Schmid, U. Diebold and C. Franchini, Physical Review B **91** (8), 085204 (2015).
57. N. A. Deskins and M. Dupuis, Physical Review B **75** (19), 195212 (2007).
58. M. Setvin, C. Franchini, X. Hao, M. Schmid, A. Janotti, M. Kaltak, C. G. Van de Walle, G. Kresse and U. Diebold, Physical Review Letters **113** (8), 086402 (2014).
59. X. Lu, P. Stepanov, W. Yang, M. Xie, M. A. Aamir, I. Das, C. Urgell, K. Watanabe, T. Taniguchi, G. Zhang, A. Bachtold, A. H. MacDonald and D. K. Efetov, Nature **574** (7780), 653-657 (2019).
60. S. Zhang, in *Handbook of Materials Modeling : Methods: Theory and Modeling*, edited by W. Andreoni and S. Yip (Springer International Publishing, Cham, 2018), pp. 1-27.
61. S. Zhang, T. Wei, J. Guan, Q. Zhu, W. Qin, W. Wang, J. Zhang, E. W. Plummer, X. Zhu, Z. Zhang and J. Guo, Physical Review Letters **122** (6), 066802 (2019).
62. E. Suárez Morell, J. D. Correa, P. Vargas, M. Pacheco and Z. Barticevic, Physical Review B **82** (12), 121407 (2010).

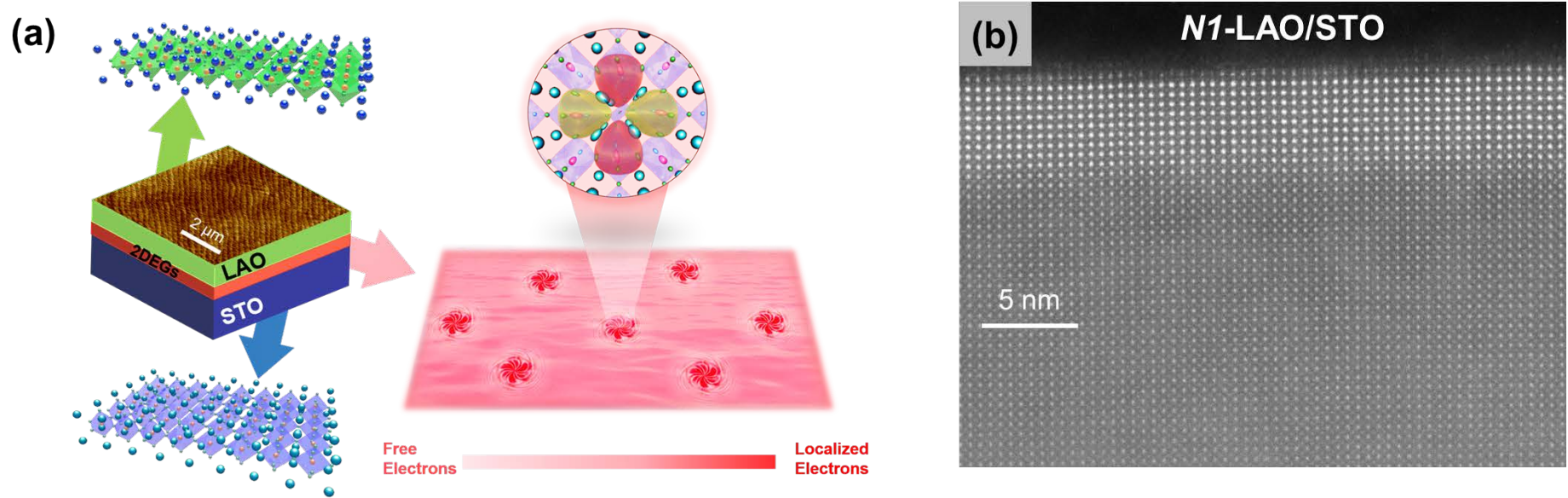

**FIG 1.** (a) Schematic displaying the formation of small polarons at the LAO/STO conducting interface with the AFM image for *N1*-LAO/STO indicating that the surfaces consist of clear atomic terraces. (b) Atomic resolution HAADF–STEM image of the representative *N1*-LAO/STO interface along the [001] zone axis. Note that the LAO thin-film is protected with Au thin-film before the preparation of TEM sample.

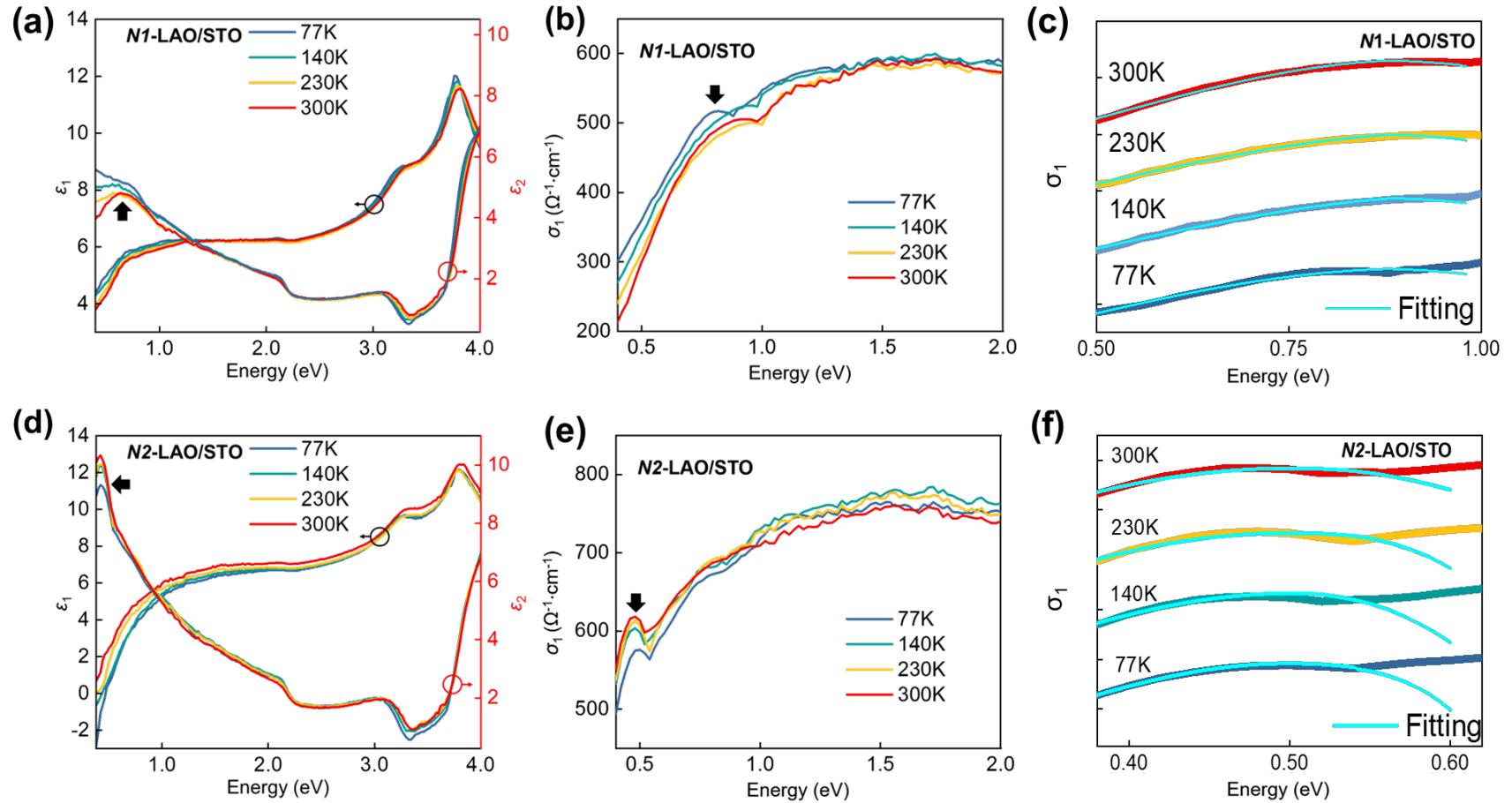

FIG 2. (a) Real, $\varepsilon_1$, and imaginary, $\varepsilon_2$, components of the dielectric function for *N1*-LAO/STO and their variations with temperature. (b) Temperature-dependent optical conductivity, $\sigma_1$, and (c) its stacked zoom-in spectra overlayed with the data fitting at the respective temperature using the Bryksin small polaron model for *N1*-LAO/STO. (d) Temperature-dependent real, $\varepsilon_1$, and imaginary, $\varepsilon_2$, components of the dielectric function for *N2*-LAO/STO. (e) Optical conductivity, $\sigma_1$, and (f) its stacked zoom-in spectra overlayed with the data fitting at the respective temperature using the Bryksin small polaron model for *N2*-LAO/STO.

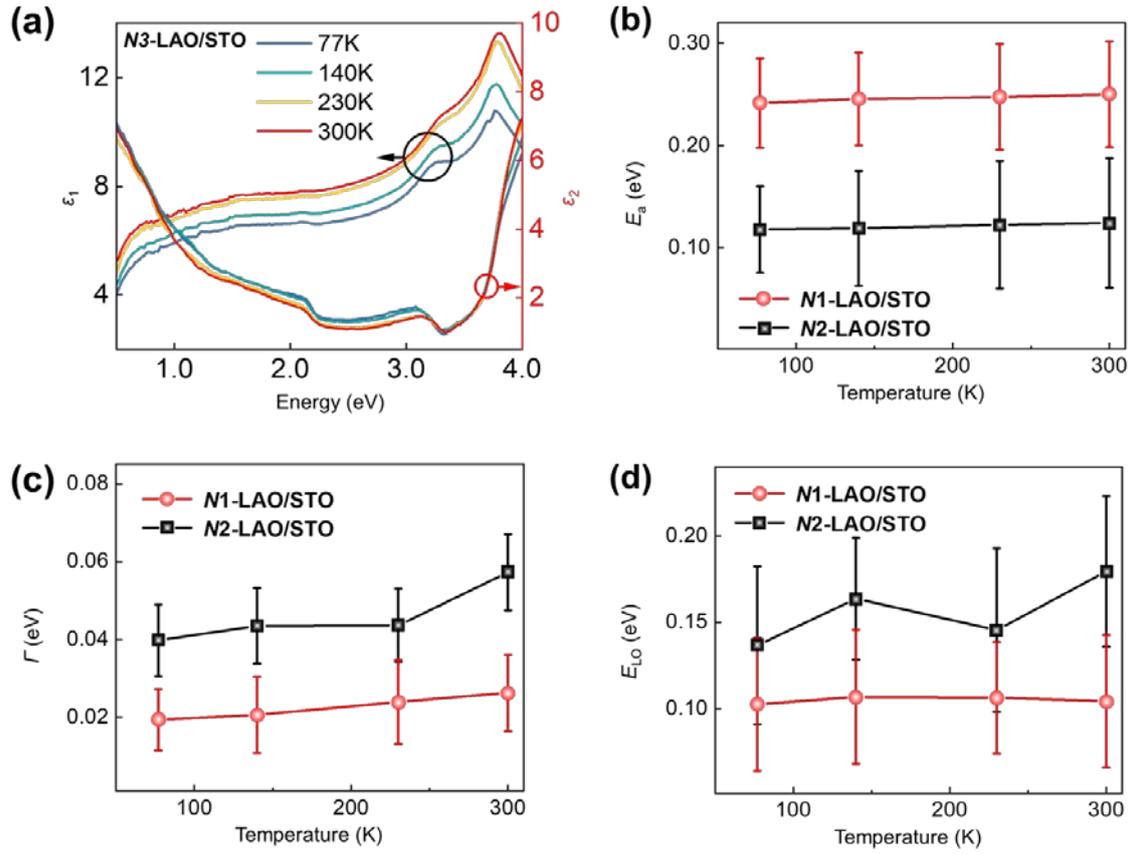

**FIG 3.** (a) Real, $\varepsilon_1$, and imaginary, $\varepsilon_2$, components of the dielectric function for the most conducting *N3*-LAO/STO interface. (b) Properties of the small polaron states of *N1*- and *N2*-LAO/STO derived using the Bryksin small-polaron model including the polaron activation or hopping energy, $E_A$. (c) The small polaron band width, $\Gamma$. (d) The phonon energy, $E_{LO}$,

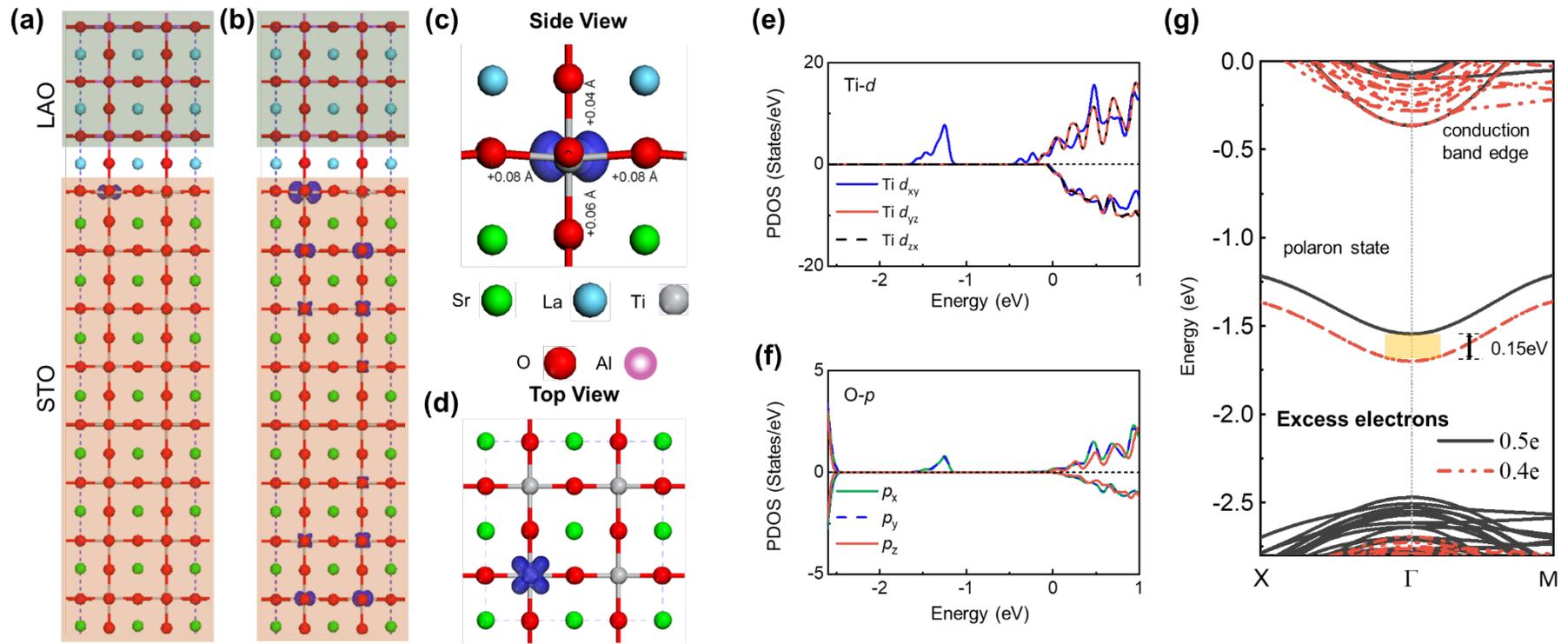

**FIG 4.** (a) The (LaAlO$_3$)$_{6.5}$/(SrTiO$_3$)$_{6.5}$ superlattice (side view) modelled for the DFT calculations. The partial charge density of the small polaron state is superimposed (blue shade) and visualized. (b) Penetration, and localization of charges at Ti lattice sites in the STO layer (blue superimposed surfaces) but localized charge concentration is sufficiently large for the small polaron formation only at the interface. (c) The side view of the LAO/STO interfacial structure. (d) The top view of the LAO/STO interfacial structure overlayed with the charge distribution of the small polaron states. The bond distortion is also denoted in the side view. (e) The Project density of states (PDOS) on the interfacial Ti-atom with the small polaron state. (f) The PDOS of the interfacial O-atom with the small polaron state. (g) The band structure (majority-spin) of (LaAlO$_3$)$_{6.5}$/(SrTiO$_3$)$_{6.5}$ superlattices with excess 0.5 electrons (black solid lines) and 0.4 electrons (red dashed lines) per unit cell. The relative shift of the small polaron states between the interfaces of different charge concentration is ~0.15 eV as highlighted by the yellow shade.